\documentclass[12pt]{iopart}
\usepackage{epsfig,float,afterpage,amssymb}
\newcommand{\be}{\begin{equation}}
\newcommand{\ee}{\end{equation}}
\newcommand{\bea}{\begin{eqnarray}}
\newcommand{\eea}{\end{eqnarray}}
\newcommand{\w}{\omega}
\newcommand{\bra}{\langle}
\newcommand{\cc}{\mbox{\scriptsize{c}}}
\newcommand{\ii}{\mbox{\scriptsize{i}}}
\newcommand{\m}{\mbox{\scriptsize{m}}}
\newcommand{\K}{\mbox{\scriptsize{K}}}
\newcommand{\ket}{\rangle}
\newcommand{\subi}{\mbox{\scriptsize{I}}}
\newcommand{\subr}{\mbox{\scriptsize{R}}}

\newcommand{\delr}{\Delta_0^{\frac{1}{1-r}}}
\newcommand{\up}{\uparrow}
\newcommand{\down}{\downarrow}
\newcommand{\ra}{\rightarrow}
\newcommand{\ssz}{\scriptsize}

\newcounter{saveeqn}
\newcommand{\alpheqn}{\setcounter{saveeqn}{\value{equation}}%
\setcounter{equation}{0}%
\addtocounter{saveeqn}{1}%
\renewcommand{\theequation}{\mbox{\arabic{section}.\arabic{saveeqn}\alph{equation}}}%
}

\newcommand{\reseteqn}{\setcounter{equation}{\value{saveeqn}}%
\renewcommand{\theequation}{\arabic{section}.\arabic{equation}}}

\newcommand{\seceq}{\setcounter{equation}{0}}

\begin{document}
\jl{31}
\title[Anderson impurities in gapless hosts]{Anderson impurities in gapless hosts: comparison of renormalization group and local moment approaches}
\author{Ralf Bulla$^{\dagger}$, Matthew T Glossop$^{\ddagger}$, David E Logan$^{\ddagger}$\\
and Thomas Pruschke$^{\S}$}
\address{$^{\dagger}$ Theoretische Physik III, Elektronische Korrelationen und Magnetismus, Universit\"{a}t Augsburg, D-86135 Augsburg, Germany}
\address{$^\ddagger$ Oxford University, Physical and Theoretical Chemistry Laboratory, South Parks Road, Oxford OX1 3QZ, UK}
\address{$^{\S}$ Institut f\"{u}r Theoretische Physik der Universit\"{a}t, 93040 Regensburg, Germany}
\tolerance=50
\begin{abstract}
 The symmetric Anderson impurity model with a hybridization
vanishing at the Fermi level, $\Delta_{\subi} \propto |\omega|^{r}$, is 
studied via the numerical renormalization group (NRG) at $T=0$; and detailed
comparison made with predictions arising from the local moment
approach (LMA), a recently developed many-body theory which is found 
to provide a remarkably successful description of the problem.
Results for the `normal' ($r=0$) impurity model are obtained as a
specific case, and likewise compared. 
Particular emphasis is given 
both to single-particle excitation dynamics, and to the transition
between the strong coupling (SC) and local moment (LM) phases of the
model. Scaling characteristics and asymptotic behaviour of the SC/LM
phase boundaries are considered. Single-particle spectra $D(\omega)$
are investigated in some detail, for the SC phase in particular. Here,
in accordance with a recently established result, the
modified spectral functions ${\cal F}(\omega) \propto |\omega|^{r}D(\omega)$
are found to contain a generalized Kondo resonance that is ubiquitously
pinned at the Fermi level; and which exhibits a characteristic low-energy
Kondo scale, $\omega_{\K}(r)$, that narrows progressively
upon approach to the SC$\rightarrow$LM transition, where it vanishes.
Universal scaling of the spectra as the transition is approached thus
results. The scaling spectrum
characteristic of the normal Anderson model is recovered as a particular
case, that exemplifies
behaviour characteristic of the SC phase generally, and which is captured
quantitatively by the LMA. In all cases the $r$-dependent scaling
spectra are found to possess characteristic low-energy 
asymptotics, but to be dominated by generalized Doniach-\u{S}unji\'{c} 
tails, in agreement with LMA predictions. 
\end{abstract}
\maketitle
\section{Introduction}
The Anderson impurity model [1], in which a correlated spin-$\frac{1}{2}$ impurity
is coupled to, and quenched by, the low-energy excitations of a non-interacting
metallic host, has long occupied
a central role in condensed matter theory (for a comprehensive review see e.g.\ [2]).
Its low-energy properties ---  those of a conventional Fermi liquid ---
are of course contingent upon coupling to a \it metallic \rm host, whose density
of states $\rho_{\mbox{\ssz{host}}}(\omega)$ is non-vanishing at the Fermi level, $\omega =0$.
However the `normal' impurity model is itself a limiting case ($r=0$) of a more
general class, in which the host spectrum contains a soft-gap at the Fermi level,
$\rho_{\mbox{\ssz{host}}}(\omega) \propto |\omega|^{r}$ with $r>0$. 
Such models, exhibiting a much wider spectrum of physical behaviour than the $r=0$
limit (as discussed below), are naturally of considerable theoretical interest; but not solely so, there being a number of potential candidates
for soft-gap behaviour, including various semiconductors [3] and heavy Fermion
superconductors [4], certain two-dimensional systems [5,6] and
quasi-one-dimensional metals described by a Luttinger model [7].

  The soft-gap problem was first considered by Withoff and Fradkin [8] in relation to
the corresponding Kondo model. Much study of both soft-gap Kondo and Anderson models
has since ensued, in particular via poor man's scaling [8,9], large-$N$ expansions
(with $N$ the impurity degeneracy) [8,10,11], the numerical renormalization group
(NRG) [9,12-15] and perturbation theory in the impurity interaction strength, $U$ [16]. 
From such, it is now well established that two distinct types of ground state 
arise, between which in general a quantum phase transition occurs at a critical,
finite host-impurity coupling: (i) a doubly degenerate local moment (LM) state
in which the impurity spin remains unquenched; and (ii) a strong coupling (SC) state
in which the impurity spin is locally quenched and a Kondo effect manifest, and whose
properties have been argued to represent a nontrivial but natural generalization
of the Fermi liquid behaviour ubiquitous to the normal Anderson model [15,16]. The
underlying physics is known to be particularly rich for the particle-hole symmetric
problem, to which NRG studies in particular have devoted considerable attention,
encompassing both thermodynamic properties [12-15] and impurity single-particle
spectra [14].

  For impurity models generally, a theoretical description of dynamics, and in
particular single-particle dynamics, poses a significant challenge. Even for the
normal spin-$\frac{1}{2}$ Anderson model, current theories such as the non-crossing
approximation [17-20], $1/N$ expansions [21-23] or slave boson approaches [24-26] are well known
to possess significant limitations; whether it be an inability in practice to handle
finite interaction strengths, to describe excitation dynamics on all energy scales,
or to recover low-energy spectral characteristics (Fermi liquid behaviour, for instance, or Doniach-\u{S}unji\'{c} tails [27] in the `wings' of the Kondo resonance). Benchmark
results are of course now accessible via NRG calculations (see e.g.\  [14,28,29]), or by a
combination of Quantum Monte Carlo calculations and the maximum-entropy method
(at finite temperature) [30]. It is against such that the accuracy, qualitative
or otherwise, of extant theories --- or new ones --- must be assessed.

  In this paper we consider the symmetric soft-gap Anderson model, including the
`normal' limit of $r=0$. Our primary aim is to make detailed comparison between
NRG calculations and a new theoretical approach to the problem, known as the local
moment approach (LMA), that has recently been developed to treat the soft-gap
Anderson model in [31] (denoted hereafter as I); and from which results for the normal
model, to which the approach was originally applied [32], are recovered smoothly
in the limit $r \rightarrow 0$.

  The dominant focus of the LMA is a treatment of single-particle
dynamics, embodied in the impurity single-particle spectrum $D(\omega)$
--- on all energy scales and for any interaction strength --- although an
essential element of it also permits the SC/LM transition, and associated
phase boundaries, to be addressed directly. The LMA is naturally 
non-perturbative, with both the notion of local moments and the possibility of
either a SC or LM state introduced explicitly and \it self-consistently \rm
from the outset, as reflected in the adoption of an underlying two-self-energy
description for the impurity Green function. The approach itself is introduced
briefly in \S2 (full details being given in I); where in addition we emphasize
the importance of a $U$-independent pinning theorem for the single-particle
spectrum of the SC phase at the Fermi level, recently established on general
grounds in [16]. We then outline a number of predictions arising from the LMA that,
in addition to single-particle spectra, can be tested directly against NRG
calculations. These include two regimes of
one-parameter scaling for the SC/LM phase boundaries; asymptotically
exact results for the small-$r$ behaviour of both the phase boundaries and
Kondo scale characteristic of the SC phase; and the prediction that, as for
the normal $r=0$ Anderson model, the so-called modified spectral functions
${\cal F}(\omega) \propto |\omega|^{r}D(\omega)$ should exhibit scaling
behaviour upon approach to the SC$\rightarrow$LM phase boundary --- where
the Kondo scale vanishes --- resulting thereby in an $r$-dependent family of universal
scaling spectra. 

  The NRG procedure employed is outlined in \S3, with particular emphasis on how
information on ground state properties can be extracted, and how they depend
upon the parameters of the NRG calculations. 
In \S's 4-6 we turn to detailed comparison between NRG and LMA results. Phase
boundaries, including their scaling characteristics and asymptotic behaviour,
 are considered in \S4; comparison with previous NRG calculations [14,15] is also
made. Single-particle spectra are considered in \S5, beginning with direct
comparison between NRG and LMA results for the `bare' $D(\omega)$, for both
SC and LM phases. This is followed by consideration of the modified spectra
${\cal F}(\omega)$, which have been argued (see I and [16]) to provide a much more
revealing description of 
single-particle dynamics characteristic of the SC phase: in direct parallel to
the normal Anderson model, the ${\cal F}(\omega)$ --- for which the pinning
theorem [16] is confirmed numerically --- are seen to contain 
a generalized Kondo resonance whose width is proportional to the Kondo
scale $\omega_{\K} \equiv \omega_{\K}(r)$,  progressively narrows as the
SC$\rightarrow$LM transition is approached, and vanishes at the transition
itself.

  The latter leads naturally to the issue of spectral scaling, to which
we turn in \S6. We begin by considering the normal Anderson model, for which
the strong coupling (large-$U$) NRG scaling spectrum $D \equiv D(\omega/\omega_{\K})$
is first obtained, and compared to the LMA result [32]. The quantitative agreement
between the two is quite remarkable; and in qualitative terms, while characteristic
Fermi liquid behaviour is naturally recovered for $|\omega|/\omega_{\K}<<1$, the
scaling spectrum is clearly dominated by the Doniach-\u{S}unji\'{c} (DS) power-law tails
[27,28,30] indicative of an incipient orthogonality catastrophe. For $r>0$, universal
scaling of the SC ${\cal F}(\omega)$ is likewise found from NRG calculations. The
$r$-dependent scaling spectra are extracted, and indeed found to possess both the
characteristic low-frequency ($|\omega|/\omega_{\K}<<1$) asymptotics and generalized
DS tails that the LMA predicts.

  Our overall conclusion from this study is that the LMA provides a rather
successful description of the soft-gap and normal Anderson models, and appears
to transcend many limitations of extant theoretical approaches. Its specific
predictions are well borne out by benchmark NRG calculations; the level of
agreement with which, both qualitative and quantitative, provides encouraging
impetus to further study and development of the basic approach.

\setcounter{section}{1}
\seceq
\section{Background}
In standard notation, and with the Fermi level taken as the origin of energy, the Hamiltonian for the spin-$\frac{1}{2}$ Anderson model is given by
\alpheqn
\reseteqn
\setcounter{equation}{0}
\begin{eqnarray}
\hat{H}& = \hat{H}_{\mbox{\ssz{host}}}+\hat{H}_{\mbox{\ssz{impurity}}}+\hat{H}_{\mbox{\ssz{hybridization}}}\nonumber \\
& = \sum_{\bi{k}, \sigma}\epsilon_{\bi{k}}\hat{n}_{\bi{k} \sigma}+\sum_{\sigma}(\epsilon_{\ii}+\mbox{$\frac{1}{2}$}U\hat{n}_{\ii-\sigma})\hat{n}_{\ii\sigma}+\sum_{\bi{k},\sigma}V_{\ii\bi{k}}(c^{\dagger}_{\ii\sigma}c_{\bi{k}\sigma}+c^{\dagger}_{\bi{k}\sigma}c_{\ii\sigma})
\eea
where $\epsilon_{\bi k}$ is the host dispersion (with corresponding spectrum $\rho_{\mbox{\ssz{host}}}(\w)$), $V_{\ii{\bi k}}$ is the hybridization matrix element and $\epsilon_{\ii}$ the impurity level; for the symmetric model considered here, $\epsilon_i=-U/2$ with $U$ the on-site Coulomb interaction.  The host-impurity coupling  is embodied succinctly in the hybridization function $\Delta(\w)=\sum_{\bi k}|V_{\ii{\bi k}}|^2/(\w+\mbox{i}\eta\mbox{sgn}(\w)-\epsilon_{\bi k})$ (where $\eta = 0+$), such that $\Delta(\w)=\Delta_{\subr}(\w)-\mbox{isgn}(\w)\Delta_{\subi}(\w)$ with
\be
\Delta_{\subi}(\w)=\pi\sum_{\bi k}|V_{\ii{\bi k}}|^2\delta(\w-\epsilon_{\bi k}).
\ee
For the soft-gap model, $\Delta_{\subi}(\w)$ takes the power-law form
\be
\Delta_{\subi}(\w)=\Delta_0|\w|^r\theta(D-|\w|)
\ee
with $r>0$ and bandwidth $D$ ($\theta(x)$ being the unit step function). Note from Eq.\ (2.2) that the form Eq.\ (2.3) arises from a constant $V_{\ii{\bi k}}$ and a soft-gap host spectrum $\rho_{\mbox{\ssz{host}}}(\w)=\sum_{\bi k}\delta(\w-\epsilon_{\bi k})$, although a separate specification of the $\{V_{\ii \bi k}\}$ and host eigenvalues $\{\epsilon_{\bi k}\}$ is not in fact required to specify $\Delta_{\subi}(\w)$.  The corresponding real part $\Delta_{\subr}(\w)$ follows by Hilbert transformation, its low-$\w$ behaviour being given by $\Delta_{\subr}(\w)=-\mbox{sgn}(\w)\mbox{tan}(\mbox{$\frac{\pi}{2}$}r)\Delta_0|\w|^r+\mbox{O}(|\w|/D)$ for $r<1$ (as here considered in practice).  The problem is thus characterized by three bare energy scales, viz $U$, $D$ and (as convenient) either $\delr$ or $\Delta_0D^r$; one of which may be used as the energy unit, resulting in two independent dimensionless parameters.

The normal Anderson model, $r=0$ is recovered as a particular limit of Eq.\  (2.3).  Here, quenching of the impurity spin is ubiquitous and the system well known (see e.g.\  [2]) to be a conventional Fermi liquid for all $U \geq 0$ and any $\Delta_0>0$ ($\Delta_0=0$ corresponding to the atomic limit where the impurity/host trivially decouple).  For $r>0$ by contrast, two distinct possible ground states are known to exist [9,14,15] --- a doubly degenerate local moment (LM) state, and a strong coupling (SC) state in which the impurity spin is locally quenched and a Kondo effect manifest --- between which in general a nontrivial quantum phase transition occurs at a critical coupling strength.  This will be investigated in the following sections, employing both the NRG and LMA.  Details of the former will be given in \S3; first, we outline the local moment approach and its predictions [31,32] relevant to the present work.

\subsection{Dynamics and the LMA}
The LMA focuses explicitly on single-particle dynamics, embodied in the zero temperature impurity Green function $G(\w)\ (\leftrightarrow G(t)=-\mbox{i}\bra T\{c_{i\sigma}(t)c^{\dagger}_{i\sigma}\}\ket)$ and hence single-particle spectrum $D(\w)=-\pi^{-1}\mbox{sgn}(\w)\mbox{Im}G(\w)$; and $G(\w)$ is usually expressed as 
\be
G(\w)=\left[\w+\mbox{i}\eta\mbox{sgn}(\w)-\Delta(\w)-\Sigma(\w)\right]^{-1}
\ee
which simply defines the conventional single self-energy $\Sigma(\w)$ [33].  As far as dynamics are concerned, our primary interest in this paper resides in the SC phase with its attendant low-energy Kondo scale.  In this regard we first note an important conservation upon the single-particle spectrum at the Fermi level, $\w = 0$, hitherto established in [16] on general grounds: namely, defining a modified spectral function $\cal{F}(\w)$ by
\be
{\cal{F}}(\w)=\pi \Delta_0\left[1+\mbox{tan}^2(\mbox{$\frac{\pi}{2}$}r)\right]|\w|^rD(\w),
\ee
that
\be
{\cal{F}}(\w=0)=1
\ee
throughout the SC phase.  For $r=0$, this recovers as a particular case a well known result for the normal impurity model, usually viewed as a consequence of the Friedel sum rule (see e.g.\  [2]): that $\pi \Delta_0 D(\w=0)=1$ --- the spectrum is always pinned at the Fermi level $\w =0$.  Equation (2.6) generalizes this pinning condition to arbitrary $r$ for a SC state and, as discussed in [16], embodies the fact that many-body interactions have no influence in renormalizing the low-$\w$ behaviour of $D(\w)$ in the SC phase:  its $\w \ra 0$ asymptotic form $D(\w)\sim|\w|^{-r}$ is precisely that of the non-interacting limit, consistent with the view [15,16,31] that the $r>0$ SC state constitutes a non-trivial but natural generalization of Fermi liquid physics.  The extent to which the pinning Eq.\  (2.6) is captured in practice should also provide a good test of the accuracy of NRG calculations, as considered in \S5.

In seeking to describe single-particle dynamics for both SC and LM phases, the LMA [31,32] eschews direct calculation of the single self-energy $\Sigma(\w)$, and instead employs a two-self-energy description with $G(\w)$ expressed formally as
\alpheqn
\be
G(\w)=\mbox{$\frac{1}{2}$}\left[G_\up(\w)+G_\down(\w)\right]
\ee
where
\be
G_\sigma(\w)=\left[\w+\mbox{i}\eta\mbox{sgn}(\w)-\Delta(\w)-\tilde{\Sigma}_\sigma(\w)\right]^{-1}
\ee
\reseteqn
(and $\sigma=\up/\down$ or $+/-$). The interaction self-energies $\tilde{\Sigma}_\sigma(\w)$ ($=-\tilde{\Sigma}_{-\sigma}(-\w)$ by particle-hole symmetry) are separated as [33]
\be
\tilde{\Sigma}_\sigma(\w)=-\mbox{$\frac{\sigma}{2}$}U|\mu|+\Sigma_\sigma(\w)
\ee
into a purely static Fock contribution (with local moment $|\mu|$) that alone would be retained at the simple mean-field level of unrestricted Hartree Fock (UHF), together with an $\w$-dependent contribution $\Sigma_\sigma(\w)$ containing the dynamics that, at low frequencies in particular, are naturally central to the problem.  There are several reasons for adopting a two-self-energy description.  First, it is a physically natural description of the doubly degenerate LM state, the self-consistent possibility of which must be introduced explicitly from the outset if one seeks to devise a non-perturbative approach that can simultaneously handle the possibility of both LM and SC states.  Second, it provides a tangible means of developing a many-body approach to the problem that starts from, but successfully transcends the limitations of, the static mean-field description (UHF).  Finally, conventional perturbation theory (PT) in $U$ about the non-interacting limit, which underpins traditional diagrammatic approaches to the single self-energy $\Sigma(\w)$, is at best limited: for example the existence of a SC/LM transition at a finite interaction strength for $0<r<\frac{1}{2}$ [9,14,15] attests to a finite radius of convergence for conventional PT, while for $\frac{1}{2}<r<1$ there is evidence to suggest a vanishing radius of convergence for PT in $U$ [16].

The LMA, described in detail in I and [32], has in practice two essential elements.  (i) It includes in the self-energies $\Sigma_\sigma(\w)$ a non-perturbative class of diagrams (Fig.\  3 of I) that embody dynamical coupling of single-particle excitations to low-energy transverse spin fluctuations, and hence capture the spin-flip scattering essential to describe the Kondo, or spin-fluctuation, regime.  Other classes of diagrams, involving primarily charge and longitudinal spin fluctuations, may also be included in the $\Sigma_\sigma(\w)$ (see e.g.\  Fig.\  9 of [32]); but retention of the dynamical spin-flip scattering processes is essential.  (ii) In describing the SC phase (and hence delineating the boundaries thereof), the spectral pinning Eq.\  (2.6) at the Fermi level $\w =0$ precisely, is enforced as a self-consistency condition; a necessary/sufficient condition for which is readily shown (\S5.1 of I) to be that ($\mbox{Re}\tilde{\Sigma}_\sigma(\w=0)\equiv$) $\tilde{\Sigma}^{\subr}_\sigma(\w=0)=0$, i.e.\  $\Sigma^{\subr}_\sigma(\w=0)=\frac{\sigma}{2}U|\mu|$ from Eq.\  (2.8).  But $\Sigma_\sigma(\w)$ depends both explicitly on $U$ and parametrically on $\frac{\sigma}{2}U|\mu|$ (since the bare propagators that enter $\Sigma_\sigma(\w)$ are of mean-field form and thus depend upon the static mean-field self-energy $-\frac{\sigma}{2} U|\mu|$).  Hence, enforcement of the $\w=0$ spectral pinning characteristic of the SC state is guaranteed by
\be
\Sigma^{\subr}_\sigma(\w=0; U; \mbox{$\frac{\sigma}{2}$}U|\mu|)=\mbox{$\frac{\sigma}{2}$}U|\mu|.
\ee

For a chosen $r$ and $U$, Eq.\  (2.9) is a self-consistency equation for the local moment $|\mu|$; the limits of stability of solutions to which, upon increasing $U$, thus yield the SC/LM phase boundary (which is thereby found, correctly, to be coterminus with that obtained upon approach from the LM phase, see \S6 of I).  Most importantly, as detailed in I, self-consistent solution of Eq.\  (2.9) introduces naturally into the problem a low energy scale that (obviously) has no counterpart at crude mean-field level: the Kondo, or spin-flip scale $\w_{\m}\equiv \w_{\m}(r)$, whose physical significance within the LMA is twofold.  First, it corresponds to the energy cost to flip the impurity spin (as manifest in a strong resonance centred upon $\w = \w_{\m}$ in the transverse spin polarization propagator [31,32]).  With increasing interaction strength in the SC phase, $\w_{\m}(r)$ diminishes progressively and vanishes as the SC/LM phase boundary is approached; while $\w_{\m}(r)=0$ throughout the LM phase --- as expected physically, since for a doubly degenerate LM state with finite weight on the impurity there is no energy cost to flip a spin.  Second, in the SC phase, $\w_{\m}(r)>0$ sets the finite timescale ($\sim 1/\w_{\m}$) for restoration of the locally broken symmetry inherent to the zeroth-order mean-field level of description, as manifest for example in the fact that for $\w/\w_{\m} \ll 1$, $\tilde{\Sigma}_\up(\w)$ and $\tilde{\Sigma}_\down(\w)$ coincide with each other (and hence with the conventional single self-energy $\Sigma(\w)$); while for the LM phase by contrast, where $\w_{\m}=0$, there is naturally no such symmetry restoration and $\tilde{\Sigma}_\up(\w) \neq \tilde{\Sigma}_\down(\w)$ even as $\w \ra 0$.

The preceeding comments are merely intended to provide a brief overview of the strategy behind the LMA, and full details are given in I.  We now highlight some predictions of the approach that will be tested against NRG calculations in \S's 4--6, beginning with the SC/LM phase boundaries.

In qualitative terms, and for any $U>0$, the LMA yields the existence {\it solely} of LM states for all $r > \frac{1}{2}$.  This agrees with extant NRG calculations [14,15] (although we note that for $U=0$ precisely the ground state is known [16] to be SC for all $0<r<1$; so for $\frac{1}{2}<r<1$ there is a SC/LM transition `at' $U=0$ itself, a fact that underlies the breakdown of conventional PT in $U$ in this $r$-regime, see [16] and I).  Moreover, for the normal Anderson model $r=0$, the LMA correctly yields a normal Fermi liquid for all $U \geq 0$ and $\Delta_0>0$ [32] --- a well known fact, but one by no means guaranteed {\it a priori } in an approximate theory.  The SC/LM phase boundary is thus confined in effect to $0<r<\frac{1}{2}$, and its $r$-dependence will in general depend upon the two independent dimensionless parameters that, as mentioned above, characterize the model: one of $\Delta_0D^r/U$ or $\delr/U$, together with $U/D$.  However the LMA predicts two distinct regimes of {\it one}-parameter scaling of the phase boundaries, according to whether $U/2 \gg D$ or $\ll D$ --- corresponding physically to an impurity level $|\epsilon_i|=\frac{1}{2}U$ that lies respectively well outside or within the host conduction band of width $D$:  (a) For $U/2 \gg D$, the phase boundary depends universally on $\Delta_0D^r/U$ alone, and not upon $U/D$; while (b) for $U/2 \ll D$ the transition line depends solely on $\Delta_0U^r/U = (\delr/U)^{1-r}$.  The latter regime is known to exist from previous NRG studies [15], the former will be investigated via the NRG in \S4.

The small $r$ behaviour of both the phase boundary, and the vanishing of the Kondo scale $\w_{\m}(r)$ as the transition is approached from the SC phase, are naturally important.  These are given respectively within the LMA by (\S6.1 of I)
\be
\left(\frac{\Delta_0\lambda^r}{U}\right)_{\cc} \stackrel{r \ra 0}{\sim}\frac{\pi}{8}r
\ee
and
\be
\w_{\m}(r) \stackrel{r \ra 0}{\sim}\lambda\left[1-\frac{U}{U_{\cc}(r)}\right]^{\frac{1}{r}}
\ee
where $\lambda = \mbox{min}[D,U/2]$.  We believe the exponent of $1/r$, and Eq.\  (2.10) for the phase boundary, to be asymptotically exact as $r \ra 0$.  The latter is seen by taking the limit $r \ra 0$ of Eq.\  (2.11), using Eq.\  (2.10), to yield $\w_{\m}(r=0)=\lambda\mbox{exp}(-\pi U/8\Delta_0)$.  This is the strong coupling (large-$U$) Kondo scale for the normal Anderson model.  The prefactor of $\lambda$ is merely an approximate high-energy cut-off, but the exponent of $-\pi U/8\Delta_0$ is exact, as known from the Bethe ansatz solution [34] and poor man's scaling (see e.g.\  [2]); and recovery of it hinges on the asymptotic validity of Eq.\  (2.10) as $r \ra 0$.  Equations (2.10,11) will likewise be compared to NRG calculations in \S4.  Note further from Eq.\  (2.10) that the condition for  mapping the symmetric Anderson model onto the corresponding Kondo model under a Schrieffer-Wolff transformation [35], viz $\Delta_0 \lambda^r/U \ll 1$, is evidently satisfied along the phase boundary line as $r \ra 0$; using Eq.\  (2.10) the critical $J_{\cc}$ for the Kondo model as $r \ra 0$ can thus be obtained and, as detailed in I, recovers precisely the result obtained by Withoff and Fradkin [8] from poor man's scaling.

One strength of the LMA is its ability to describe single-particle dynamics, which are compared to NRG calculations in \S's 5 and 6.  The known low-$\w$ spectral signatures of the LM and SC phases, viz $D(\w)\sim |\w|^r$ and $|\w|^{-r}$ respectively as $\w \ra 0$ [14], are correctly recovered by the approach; and it gives excellent agreement with extant NRG results for $D(\w)$ [14] (see Fig.\  15 of I).  However, as shown in [16] and I, the most revealing expos\'{e} of single-particle dynamics in the SC phase is evident not in $D(\w)$ itself, but in the modified spectral function ${\cal{F}}(\w)\propto |\w|^rD(\w)$ (Eq.\  (2.5)) wherein the unrenormalized $|\w|^{-r}$ divergence in $D(\w)$ has been removed.  In thus `exposing' the low-frequency many-body renormalization characteristic of the Kondo effect, the Kondo scale is now directly manifest in $\cal{F}(\w)$: in addition to spectral pinning at the Fermi level $\w = 0$ (Eq.\  (2.6)), $\cal{F}(\w)$ contains a generalized Kondo resonance with a width proportional to $\w_{\m}(r)$, that narrows upon progressive increase of $U$ in the SC phase and vanishes at the SC/LM phase boundary; see \S8 of I.  This behaviour, characteristic of the SC phase for all $r<\frac{1}{2}$, is in obvious parallel to that for the normal Anderson model, and is further support for the notion of the SC phase as a `generalized Fermi liquid'.  NRG results for $\cal{F}(\w)$ will be given in \S5.

Most significantly, since the width of the Kondo resonance in $\cal{F}(\w)$ vanishes as the SC$\ra$LM transition is approached ($U \ra U_{\cc}(r)-$), the above comments suggest --- and the LMA indeed predicts (see \S8 of I) --- that $\cal{F}(\w)$ should exhibit scaling behaviour as the phase boundary is approached from the SC side: i.e.\  $\cal{F}(\w)$ should be a universal function of $\w/\w_{\m}(r)$ for any $r$ where a SC state exists.  An $r$-dependent family of universal scaling spectra is thus expected to arise, encompassing as a specific case the well known scaling for the normal Anderson model $r=0$ as $U \ra \infty$ $(\equiv U_{\cc}(r=0)$, see Eq.\  (2.10)). This is considered in \S6 where we first compare the NRG and LMA scaling spectra for the normal Anderson model.  Further, the LMA predicts the universal $\cal{F}(\w)$'s to exhibit characteristic $r$-dependent low-frequency behaviour (for $\w/\w_{\m} \ll 1$), as well as Doniach-\u{S}unji\'{c} tails [27] that are a reflection of the orthogonality catastrophe and represent a generalization of those known to arise in the $r=0$ Anderson model [28,30]; this too is compared to NRG calculations in \S6.

\section{NRG approach}
Detailed discussion of how the NRG can be applied to the soft-gap
Anderson model can be found in [9,12-15]. Here we mention briefly only
those aspects of the approach necessary to understand how e.g.\  
information on ground state properties can be extracted, and how they
depend upon the parameters of the NRG calculations.
 
 The NRG is based quite generally on a logarithmic discretization of the
energy axis, i.e.\  one introduces a parameter $\Lambda$ and divides the
energy axis into intervals $[\Lambda^{-(n+1)}, \Lambda^{-n}]$ for
$n=0,1, ...., \infty$ [36,37]. With some further manipulations [36,37] the original
model may be mapped onto a semi-infinite chain, which can then be solved
iteratively by starting from the impurity and successively adding chain sites.
The coupling between two adjacent such sites $n$ and $n+1$ vanishes as
$\Lambda^{-n/2}$ for large $n$, whence the low-energy states of the chain
with $n+1$ sites are generally determined by a comparatively small number
$N_{\mbox{\ssz{s}}}$ of states close to the ground state of the $n$-site system. In practice
one retains only these $N_{\mbox{\ssz{s}}}$ states from the $n$-site chain to set up the
Hilbert space for the $n+1$ site chain, thus preventing the usual exponential
growth of the Hilbert space as $n$ increases. Eventually, after $n_{\mbox{\ssz{NRG}}}$
sites have been included in the calculation, addition of another site will
not change significantly the spectrum of many-particle excitations; the spectrum is very close to that of the fixed point, and the calculation may be terminated.

\begin{figure}[htb]
\begin{center}
\epsfxsize=3.5in
  \epsffile{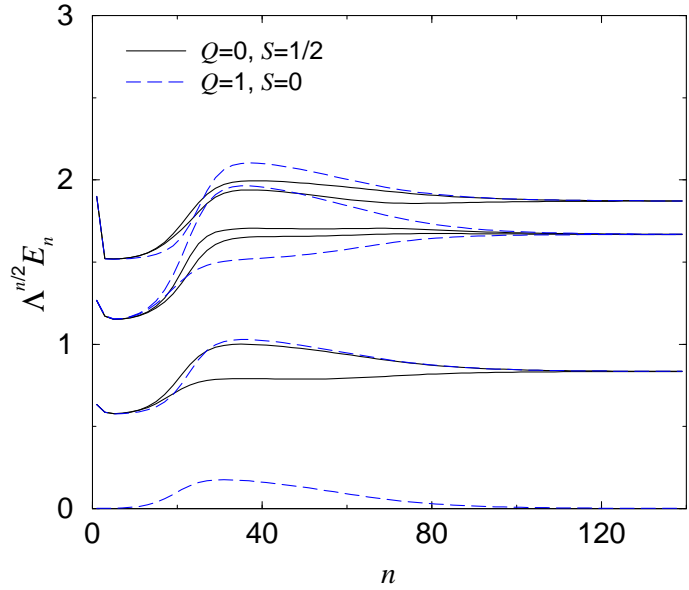}
\caption[]{NRG flow diagram for the lowest lying energy levels for
$\Delta_0D^r/D=0.0075$.  The system flows to the SC fixed point.}
\label{fig:flowSC}
\end{center}
\end{figure}
\begin{figure}[htb]
\begin{center}
\epsfxsize=3.5in
  \epsffile{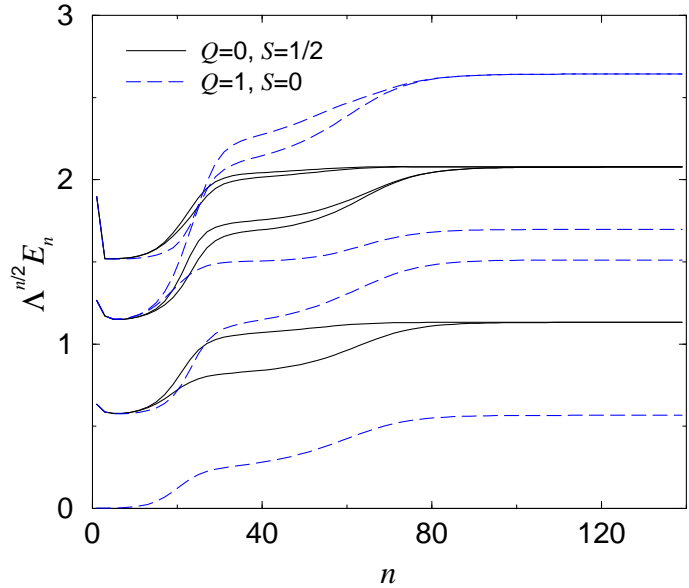}
\caption[]{\small Flow diagram for the lowest lying energy levels for
$\Delta_0D^r/D = 0.006$. The system flows to the LM fixed point.}
\label{fig:flowLM}
\end{center}
\end{figure}
  
An example of such a flow diagram for some of the lowest lying energy levels
as a function of the chain length, $n$, is given in Fig.\ 1 for the soft-gap
Anderson model with $r=0.4$, $U/D = 10^{-3}$ and $\Delta_{0}D^{r}/D = 0.0075$; 
the parameters used for the NRG calculations are $\Lambda=2$ and $N_{\mbox{\ssz{s}}} =300$.
The states are labelled by the quantum numbers $Q$ (which characterizes the number
of particles measured relative to one particle per site), and the total spin, $S$. 
As mentioned above, the energy scale is reduced in each step by a factor $\Lambda^{1/2}$.
To allow for a direct comparison of the energies for different chain lengths,
it is thus convenient to plot $\Lambda^{n/2}E_{n}$ instead of the eigenvalues
$E_{n}$ of the $n$-site chain directly. As is apparent from Fig.\ 1, the properties
of the system in this case do not change further for chain lengths $n_{\mbox{\ssz{NRG}}}>120$.
Without going into details here, one can show that the distribution of energy
levels for $n>120$ in Fig.\ 1 is characteristic of the SC phase of the model.

  If by contrast we choose instead a value of $\Delta_{0}D^{r}/D=0.006$, we obtain
the flow diagram shown in Fig.\ 2. Here it is evident that the fixed point level
structure is entirely different from the SC solution, and indeed this particular
pattern is now characteristic of the LM phase of the model. We can thus conclude,
simply from inspection of the two flow diagrams, that the critical
$(\Delta_{0}D^{r}/D)_{\cc}$ separating the SC and LM phases of the soft-gap
Anderson model for the model parameters specified, lies in the interval
$[0.006, 0.0075]$. Performing a whole series of such NRG calculations leads
to Fig.\ 3 (illustrated for the first excited state with Q=1 and S=0), and
the conclusion that
the critical $\Delta_{0}D^{r}/D$ lies between 0.00670 and 0.00675. This method
thus enables in principle a determination of the critical coupling to arbitrary
precision. Note further from Fig.\ 3 that, in the vicinity of the critical coupling, the approach to the SC or LM value of $\Lambda^{n/2}E_n$ occurs at progressively higher values of $n$.  
\begin{figure}[htb]
\begin{center}
\epsfxsize=3.5in
  \epsffile{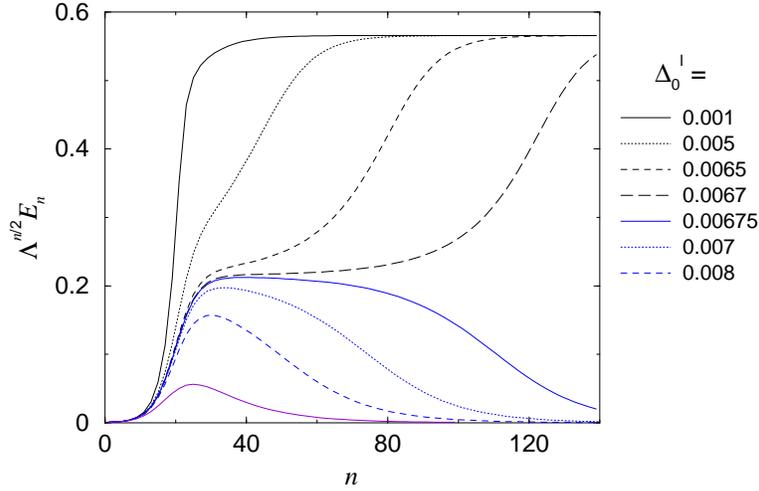}
\caption[]{\small Flow diagram for the first excited state with quantum
numbers $Q\!=\!1$ and $S\!=\!0$, for various values of
$\Delta_0'=\Delta_0D^r/D$.}
\label{fig:flowQ1S0}
\end{center}
\end{figure}
\begin{figure}[htb]
\begin{center}
\epsfxsize=3.5in
  \epsffile{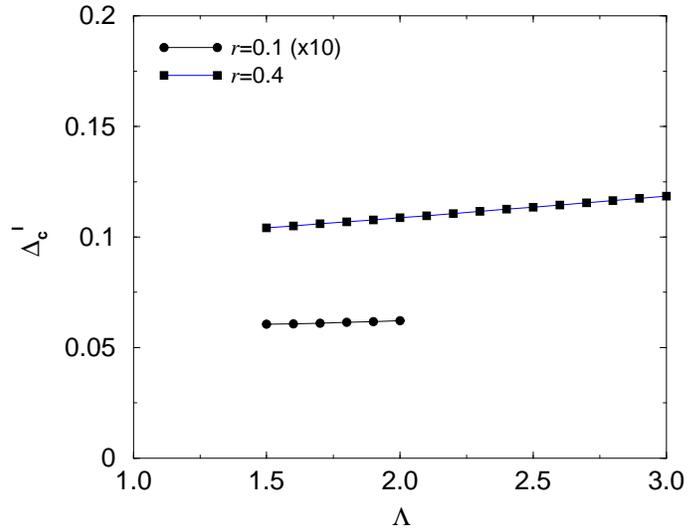}
\caption[]{\small
$\Lambda$-dependence of the critical $\Delta_{\cc}'=(\Delta_0D^r/D)_{\cc}$ for
$U/D=\!0.1$ and $r\!=\!0.1$ (circles) and 
$r\!=\!0.4$ (squares). The $\Delta_{\cc}'$
for $r\!=\!0.1$ is multiplied by a factor of 10.
}
\label{fig:Lambda-dep}
\end{center}
\end{figure}
This corresponds to the vanishing of a low-energy scale as the transition is approached from either side, the low-energy scale being proportional to the Kondo scale when coming from the SC phase.  A low-energy scale may be defined as $\w_*=\Lambda^{-n_*/2}$ with $n_*$ given by $|x_{n_*}-x_{\infty}|=0.01$, with $x_n=\Lambda^{n/2}E_n$.  In the SC phase the low-energy scale can also be determined via e.g.\ \ the width  $\w_{\K}$ of the Kondo resonance (see \S5), and the NRG results show that $\w_* \propto \w_{\K}$ (the two are not of course strictly equal, since a low-energy scale is determined only up to a prefactor).

It is of course obvious that for any $\Lambda>1$ the NRG constitutes an
approximation to the continuum system, but becomes exact in the limit
$\Lambda \rightarrow 1$. Performing this limit directly is naturally not
possible since one must simultaneously increase the number of retained
states, $N_{\mbox{\ssz{s}}}$,  to infinity; but one can study the $\Lambda$ and
$N_{\mbox{\ssz{s}}}$-dependences of the NRG results and perform the limits $\Lambda \rightarrow 1$,
$N_{\mbox{\ssz{s}}} \rightarrow \infty$ by extrapolation. 
As an example, Fig.\ 4 shows results
for the $\Lambda$-dependence of the critical $(\Delta_{0}D^{r}/D)_{\cc}$
for a fixed $N_{\mbox{\ssz{s}}} =300$; for both $r=0.1$ and $r=0.4$, and with $U/D= 0.1$.
It is evident from Fig.\ 4 that the $\Lambda$-dependence is rather mild, although
increasing slightly as $r \rightarrow \frac{1}{2}$. Similar study of the $N_{\mbox{\ssz{s}}}$-dependence
also reveals a rather weak variation that again does not depend significantly
on $r$. From such considerations we conclude that for practical purposes, a choice
of $\Lambda =2$ and $N_{\mbox{\ssz{s}}} =300$ will in general be sufficient to determine e.g.\ 
the SC/LM phase boundaries to good accuracy and with minimal numerical effort.
One should nonetheless keep in mind that due to the $\Lambda$-dependence the actual
critical coupling $(\Delta_{0}D^{r}/D)_{c}$ will always be a few percent smaller,
the deviation increasing mildly as $r \rightarrow \frac{1}{2}$.

  Finally, we remind the reader that the NRG also enables calculation of dynamical
properties such as the single-particle spectrum $D(\omega)$, as detailed in [14]
and [38]; and results for which will be given in \S's 5 and 6.

\seceq
\section{Phase boundaries}

In discussing phase boundaries between SC and LM states, we consider first the case $U/D \ll 1$ that has been the focus of previous NRG studies [14,15].  Here the phase boundary for given $r$ depends solely on the ratio $\Delta_0U^r/U$, a result known to arise from application of poor man's scaling to the soft-gap Anderson model [9].  It may however be understood quite simply by noting that $U/D \ra 0$ corresponds to the conventional limit of an infinite host bandwidth, $D \ra \infty$ (which may be taken with impunity for $r<1$, see e.g.\ I); in which limit the $D$-scale drops out of the problem, and the model thus depends solely on the dimensionless ratio $\tilde{U}=U/\delr$.  
\begin{figure}[htb]
\begin{center}
\epsfxsize=3.5in
  \epsffile{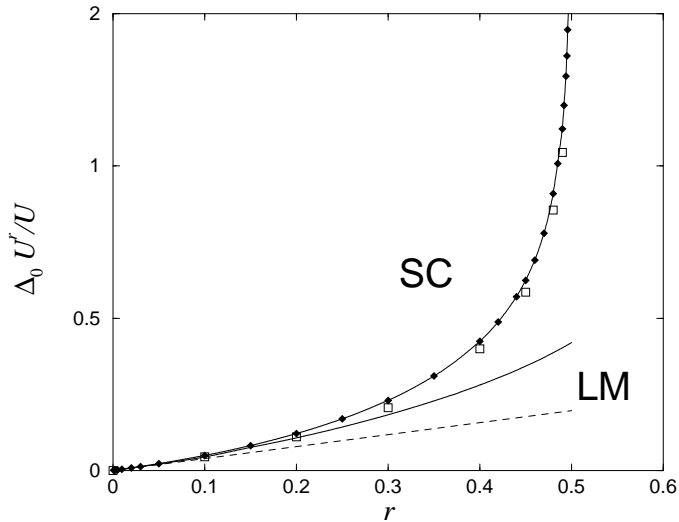}
\caption{For $U/D \ll 1$, comparison of NRG phase boundary $(\Delta_0U^r/U)_{\cc}$ versus $r$ (solid diamonds) with that obtained via the LMA (solid line).  Previous NRG results [15] are also shown (open squares), together with the $r \ra 0$ Kondo asymptote of $\pi r/8$ (dashed line).}
\label{fig5}
\end{center}
\end{figure}
\begin{figure}[htb]
\begin{center}
\epsfxsize=3.5in
  \epsffile{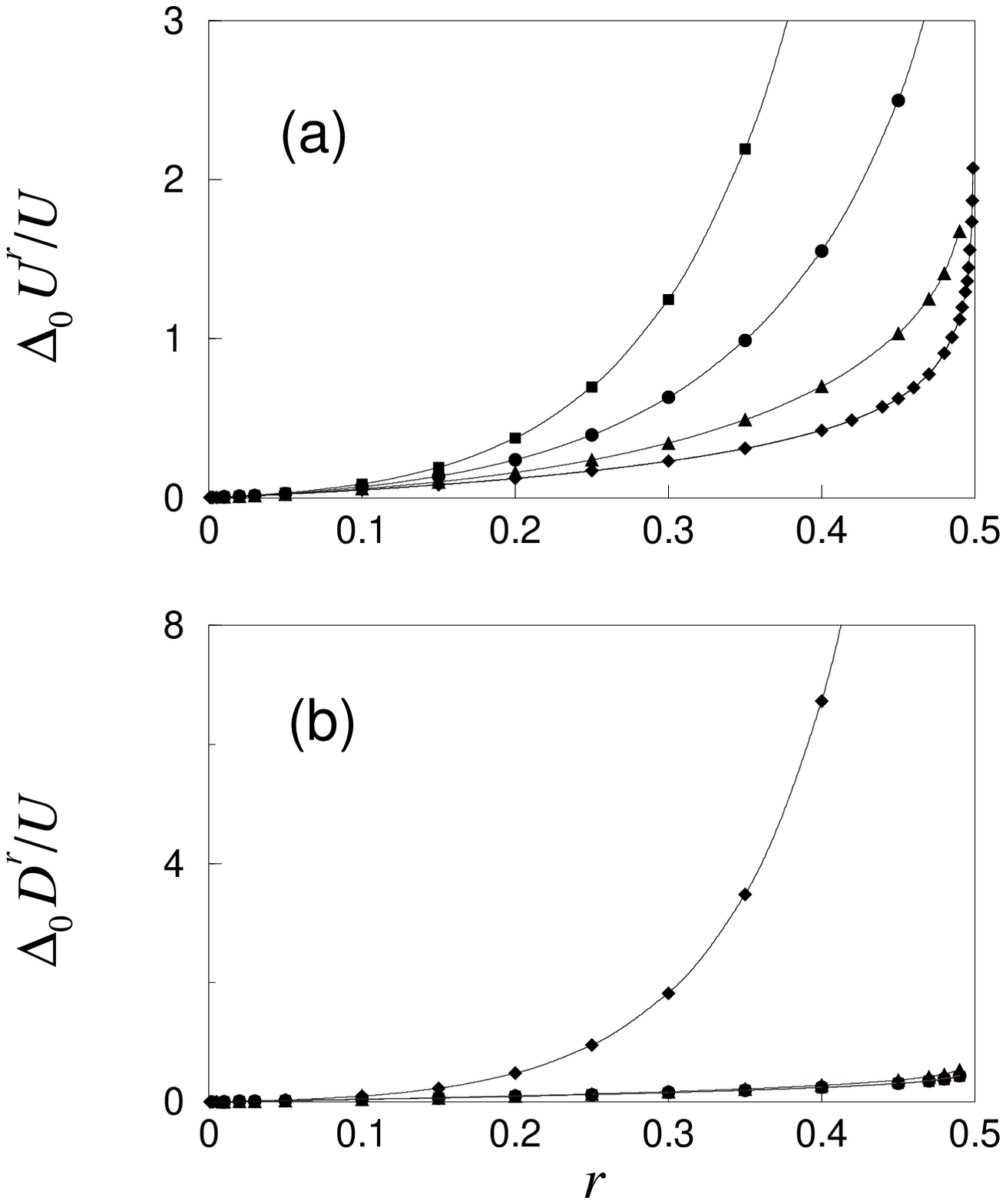}
\caption{NRG phase boundaries for $U/D$ = $10^3$ (squares), $10^2$ (circles), 10 (triangles) and $10^{-3}$ (diamonds).  Expressed as (a) $(\Delta_0U^r/U)_{\cc}$ the curves are strongly $U/D$-dependent; but as (b) $(\Delta_0D^r/U)_{\cc}$, the predicted one-parameter scaling is evident for $U/D \gg  1$.  Lines connecting points are a guide to the eye only.}
\label{fi}
\end{center}
\end{figure}  
Phase boundaries correspond to a critical $\tilde{U}_{\cc}(r)$, an alternative but equivalent way of displaying which is via $(\Delta_0U^r/U)_{\cc}=\tilde{U}_{\cc}^{r-1}(r)$.

The above behaviour for $U/D \ll 1$ has been observed in NRG calculations by Gonzalez-Buxton and Ingersent (G-BI) [15], who find in practice excellent scaling of the phase boundary for $U/D = 0.2$ and $0.02$ (see Fig.\  20 of [15], where the quantity $\rho_0J_{\cc}$ there plotted for the symmetric model is $(8/\pi)(\Delta_0\lambda^r/U)_{\cc}$ with $\lambda=U/2$).  As noted in I however, the phase boundary $(\Delta_0U^r/U)_{\cc}$ obtained originally by Bulla, Pruschke and Hewson [14] for $U/D = 10^{-3}$ differs significantly from that found by G-BI (see Fig.\  9 of I for an explicit comparison).  We have thus repeated a determination of the phase boundary for $U/D = 10^{-3}$ using the more refined procedure outlined in \S3.  As shown in Fig.\  5, good agreement with the G-BI results is now found: the difference between the two sets of data is small, and readily attributable to the choice of $\Lambda=2$ and $N_{\mbox{\ssz{s}}}=300$ in the present calculations, which (see \S3) will yield a slight overestimate of $(\Delta_0U^r/U)_{\cc}$ (the G-BI results [15] having already been extrapolated to the continuum limit).

Fig.\  5 will be discussed further below, before which we consider the opposite limit of $U/D \gg 1$.  Here, in contrast to $U/D \ll 1$, the LMA predicts (see \S6.3 of I) that none of the bare scales $U, D$ or $\delr$ drop out of the problem, but that the phase boundary exhibits one-parameter scaling when expressed in terms of $\Delta_0D^r/U (=(U/\delr)^{r-1}(D/U)^r)$.  This too is physically understandable, since for $|\epsilon_i|=U/2 \gg D$ where the impurity level lies well outside the host band, the impurity-host coupling is controlled by the hybridization $\Delta_{\subi}(D)=\Delta_0D^r$ which, together with $U$, sets the natural energy scales upon which ratio the phase boundary thus depends.  This is confirmed by NRG calculations, as shown in Fig.\  6.
In Fig.\  6a we show the critical $(\Delta_0U^r/U)_{\cc}$ versus $r$ for $U/D=10^3, 10^2, 10$ and (as in Fig.\  5) $10^{-3}$; expressed in this form, the phase boundary curves depend strongly upon the chosen $U/D$.  For the same $U/D$ ratios, Fig.\  6b by contrast shows the critical $(\Delta_0D^r/U)_{\cc}$ versus $r$, from which the predicted one-parameter scaling for $U/D \gg 1$ is seen clearly; in particular, the results for $U/D = 10^2$ and $10^3$ are essentially indistinguishable.

In Fig.\  7, NRG results for $(\Delta_0D^r/U)_{\cc}$ versus $r$ (with $U/D =10^3$) are compared to those arising from the LMA for $U/D \gg 1$; while for $U/D \ll 1$, Fig.\  5 likewise shows the corresponding LMA result for $(\Delta_0D^r/U)_{\cc}$.  In both Figs.\  5 and 7 we also indicate the predicted $r \ra 0$ asymptotic behaviour of the phase boundary that is symptomatic of the Kondo limit, viz $\pi r/8$ (Eq.\  (2.10); behaviour that is indeed recovered from the NRG calculations, regardless of whether $U/D \gg 1$ or $\ll 1$.  In the latter case, the small-$r$ behaviour is reached in practice for $r \lesssim 0.02$ (Fig.\  5); while for $U/D \gg 1$ Fig.\  7 shows that the critical $\Delta_0D^r/U$ remains close to its Kondo asymptote of $\pi r/8$ over a wider $r$-range (up to $r \sim 0.1$), as one expects physically for $U/D \gg 1$ where charge fluctuations are relatively less significant.  Note however that, even for $U/D \ra \infty$, the phase boundaries of the soft-gap Anderson and Kondo models coincide strictly only as $r \ra 0$: the condition for mapping the former onto the latter via a Schrieffer-Wolff transformation --- and hence the suppression of charge fluctuations in the Anderson model --- is $\Delta_0\lambda^r/U \ll 1$ (with $\lambda = \mbox{min}[D,U/2]$), and from Eq.\  (2.10) is satisfied asymptotically on the phase boundary as $r \ra 0$.  Charge fluctuations in the Anderson model cannot therefore be neglected entirely even as $U/D \ra \infty$, save for $r \ra 0$; as manifest in the fact (Fig.\  7) that $(\Delta_0D^r/U)_{\cc}$ remains in general finite at the SC/LM transition.

\begin{figure}[htb]
\begin{center}
\epsfxsize=3.5in
  \epsffile{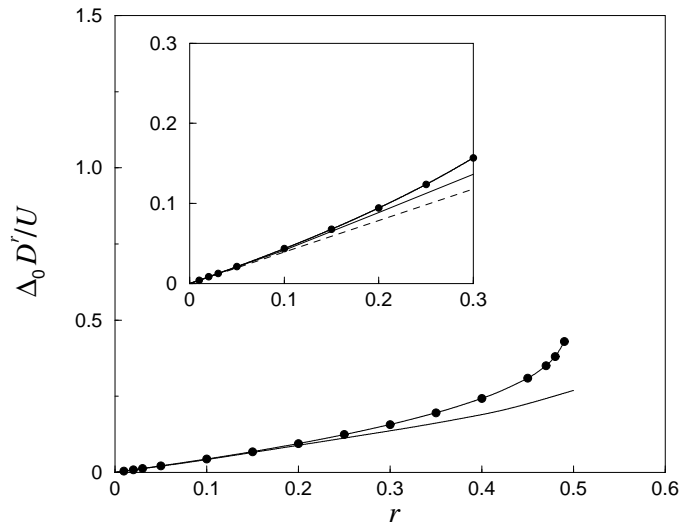}
\caption[]{NRG phase boundary $(\Delta_0D^r/U)_{\cc}$ versus r for $U/D=10^3$ (solid circles), compared to that obtained via the LMA for $U/D \gg 1$ (solid line).  Lines connecting NRG points are a guide to the eye only.  Inset: results on an expanded scale, including the Kondo asymptote of $\pi r/8$ (dashed line).}
\label{fig9}
\end{center}
\end{figure}

More broadly, Figs.\  5 and 7 show the predicted LMA phase boundaries to be in very good quantitative accord with the NRG results for $r \lesssim 0.3$ or so; although with further increasing $r$ the NRG boundaries increase more rapidly than their LMA counterparts.  This brings us to the issue of how the phase boundaries behave as $r \ra \frac{1}{2}-$ (for $r>\frac{1}{2}$, both the NRG and LMA yield solely LM states for $U>0$).  Previous NRG studies for the soft-gap Anderson model [14,15] have suggested a divergent phase boundary as $r \ra \frac {1}{2}-$, although it has been argued in I that the data themselves do not reliably warrant such a conclusion.  We have therefore re-investigated the matter more comprehensively, and conclude that the phase boundaries are indeed divergent as $r \ra \frac{1}{2}-$.  An incipient divergence is evident in the NRG results for $U/D = 10^{-3}$ shown in Fig.\  5, which has been extended up to $r=0.499$; it is relatively weak, and fits well to logarithmic behaviour, $(\Delta_0U^r/U)_{\cc} \propto -\mbox{ln}(\frac{1}{2}-r)$.  Perhaps the most compelling evidence is however obtained by fixing $r=\frac{1}{2}$ precisely, and (for the chosen $U/D$ and with $D=1$ as the energy unit) progressively increasing the hybridization $\Delta_0$: up to $\Delta_0$ values that are more than two orders of magnitude in excess of the critical $\Delta_0$ for $r=0.499$, only LM states are observed and there is no hint of a SC state at $r=\frac{1}{2}$.  Such behaviour is naturally not specific to $U/D=10^{-3}$ (although note that in Fig.\  7 for $U/D=10^3$, the final $r<\frac{1}{2}$ NRG point is $r=0.49$).

The above behaviour as $r \ra \frac{1}{2}-$ is in contrast to that found within the LMA, where the phase boundary terminates at a finite value (although note ironically that a divergent phase boundary as $r \ra \frac{1}{2}-$ arises at the simple static mean-field level, see e.g.\ Fig.\  9 of I).  This discrepancy is plausibly explained by noting that while charge fluctuations are partially included in the present (naturally approximate) LMA, it deliberately focuses on and successfully captures the strong coupling physics of the Kondo/spin-fluctuation regime that is asymptotically dominant for small-$r$.  It does not therefore do justice to the charge fluctuations that, granted a divergent phase boundary, must become increasingly important as $r \ra \frac{1}{2}-$. Incorporation of such effects into the basic approach (beyond those already present) should however be possible, and is currently under investigation.  We also add that a finite critical $J_{\cc}(r)$ as $r \ra \frac{1}{2}-$ has hitherto been reported in an NRG study of the symmetric soft-gap Kondo model [13].  This is now thought to be an artifact, however, and $J_{\cc}(r)$ is likewise believed to diverge as $r \ra \frac{1}{2}-$ [39].

Finally, we consider briefly the exponent for the vanishing of the Kondo scale as the SC/LM transition is approached from the SC phase, as reflected in the behaviour of the low-energy NRG scale $\w_*(r)$ (\S3) whose critical behaviour may be expressed as
\be
\w_*(r) \propto \left(1-\frac{p_{\cc}(r)}{p}\right)^{x(r)}.
\ee
Here, $p=\Delta_0\lambda^r/U$ (with $\lambda = \mbox{min}[D,U/2]$), and the exponent $x(r)$ is thus defined.  The LMA result for $x(r)$ is $x(r) \propto 1/r$ as $r \ra 0$ (Eq.\  (2.11)) and, while this can be established analytically only as $r \ra 0$, numerical analysis shows the exponent $1/r$ to hold generally within the LMA (\S6.1 of I).  NRG results for $x(r)$ are given in Table I; and as $r$ tends to either 0 or $\frac{1}{2}$ we estimate a numerical uncertainty of O(1) in the quoted values.  (The data shown refer specifically to $U/D=10$, although we do not believe them to be $U/D$-dependent: the same values are obtained to within numerical accuracy for $U/D=0.1$.)  

\vspace{0.5cm}
\begin{tabular}{|l||llllllllll|}
\multicolumn{11}{c}{{\bf Table I.} NRG results for the exponent $x(r)$.}\\
\hline
$r$&0.05&0.1&0.15&0.2&0.25&0.3&0.35&0.4&0.45&0.47\\ \hline
$x(r)$&19.6&10.8&7.6&6.2&5.43&5.14&5.2&5.85&8.55&12.2\\ \hline

\end{tabular}
\vspace{0.5cm}

\noindent From Table I it is apparent that $x(r)\sim 1/r$ as $r \ra 0$ is indeed recovered.  With increasing $r$, however, $x(r)$ falls off less rapidly than $1/r$, reaches a minimum at $r\sim 0.3$ and increases again as $r \ra \frac{1}{2}$ where we suspect it diverges.  While we cannot offer an explanation for this behaviour, save as $r \ra 0$, we do not doubt at least its qualitative accuracy; and in view of the comments made above it is hardly surprising that the NRG and LMA exponents differ with increasing $r$.

\seceq
\section{Single-particle dynamics}
We begin with a brief overview of spectral characteristics in both SC and LM phases for a representative $r<\frac{1}{2}$, and compare directly NRG and LMA determined spectra; $r=0.2$ is chosen, together with a fixed ratio $U/D=0.1$.  Fig.\  8a shows NRG results for the single particle spectrum versus $\tilde{\w}=\w/\delr$, for four reduced interaction strengths $\tilde{U}=U/\delr$: $\tilde{U}=$ 31.6 (LM phase), 19.0 (LM), 13.3 (SC phase) and 10.1 (SC); the SC/LM transition occurs at $\tilde{U}_{\cc} \simeq 13.7$.  Corresponding spectra obtained from the LMA for the same parameters are shown in Fig.\  8b.  

\begin{figure}[htb]
\begin{center}
\epsfxsize=3.5in
  \epsffile{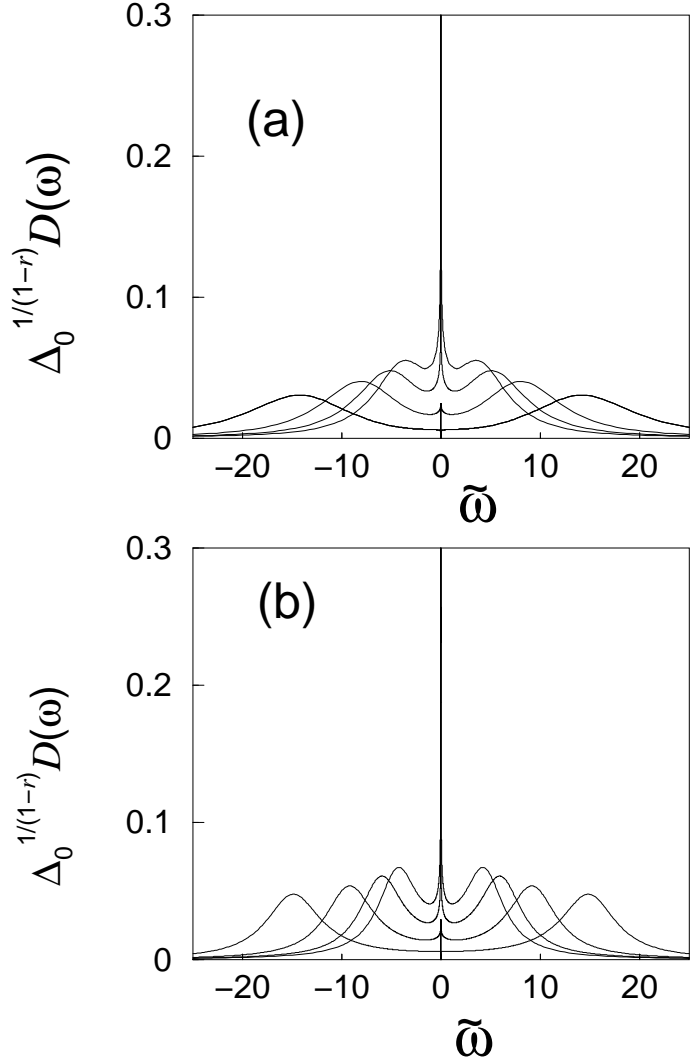}
\caption{$\delr D(\w)$ versus $\tilde{\w}=\w/\delr$. (a) NRG with fixed $r=0.2$ and $U/D=0.1$, and (from outside to inside) for: $\tilde{U}=U/\delr$ = 31.6 (LM), 19.0 (LM), 13.3 (SC) and 10.1 (SC); the critical $\tilde{U}_{\cc}=13.7$.  (b) LMA spectra for the same parameters.}
\end{center}
\end{figure}

With decreasing interaction strength in the LM phase the Hubbard satellites, centred on $\tilde{\w}= \pm \frac{1}{2}\tilde{U}$, naturally move to lower energies; and in addition a narrow low-energy feature is seen to develop in the vicinity of the Fermi energy $\w=0$.  This becomes increasingly pronounced as $\tilde{U}$ decreases toward the $\tilde{U}_{\cc}(r)$ at which the LM/SC transition occurs but, as seen from Fig.\  8, $D(\w)$ vanishes as $\w \ra 0$ in the LM phase, with the asymptotic behaviour $D(\w)\sim |\w|^r$ (behaviour that is characteristic of the LM state for any $r>0$).  For $\tilde{U}<\tilde{U}_{\cc}(r)$ by contrast, the spectra exhibit the characteristic $\w \ra 0$ divergence $D(\w)\sim |\w|^{-r}$ that is symptomatic of the SC phase for any $r<\frac{1}{2}$.  And with further decreasing interaction strength in the SC phase the Hubbard satellites naturally become less distinct, and as $\tilde{U}\ra 0$ are subsumed into the central divergent resonance (see e.g.\ Fig.\  11c of I).

As is evident from Fig.\  8, agreement between the NRG and LMA spectra is rather good (see also Fig.\  15 of I); the principal difference being that the NRG Hubbard satellites in Fig.\  8a are more diffuse than their LMA counterparts, and become increasingly so with increasing $\tilde{U}$.  However this is likely to be an artifact of the NRG calculations.  The logarithmic discretization employed in the NRG is designed to capture the low-energy physics; and on higher frequency scales appropriate to the Hubbard satellites, concomitant broadening of the discrete and relatively sparse NRG poles is well known to produce spectral overbroadening.  We add moreover that the Hubbard satellites within the LMA already contain the effects of additional many body broadening beyond simple mean-field level, well known to exist for the normal $r=0$ Anderson model (see e.g.\ [20,32]), and shown in \S7 of I to arise also for the soft-gap model; and on these grounds too the diffuse character of the NRG Hubbard satellites is unlikely to be physically correct.  

As discussed in \S2 (see also I and [16]), the modified spectral function ${\cal F}(\w)=\pi\Delta_0\left[1+\mbox{tan}^2(\frac{\pi}{2}r)\right]|\w|^rD(\w)$, which removes the $|\w|^{-r}$ divergence that is symptomatic of the SC state but unrenormalized by interaction effects, provides a much more acute probe of single-particle dynamics in the SC phase.  This is illustrated in Fig.\  9, again for $r=0.2$ and $U/D=0.1$, where NRG results for ${\cal F}(\w)$ versus $\tilde{\w}=\w/\delr$ are shown for reduced interaction strengths: $\tilde{U}$ = 8.0 (SC), 10.1 (SC), 13.3 (SC) and 14.2 (LM).

\begin{figure}[htb]
\begin{center}
\epsfxsize=3.5in
  \epsffile{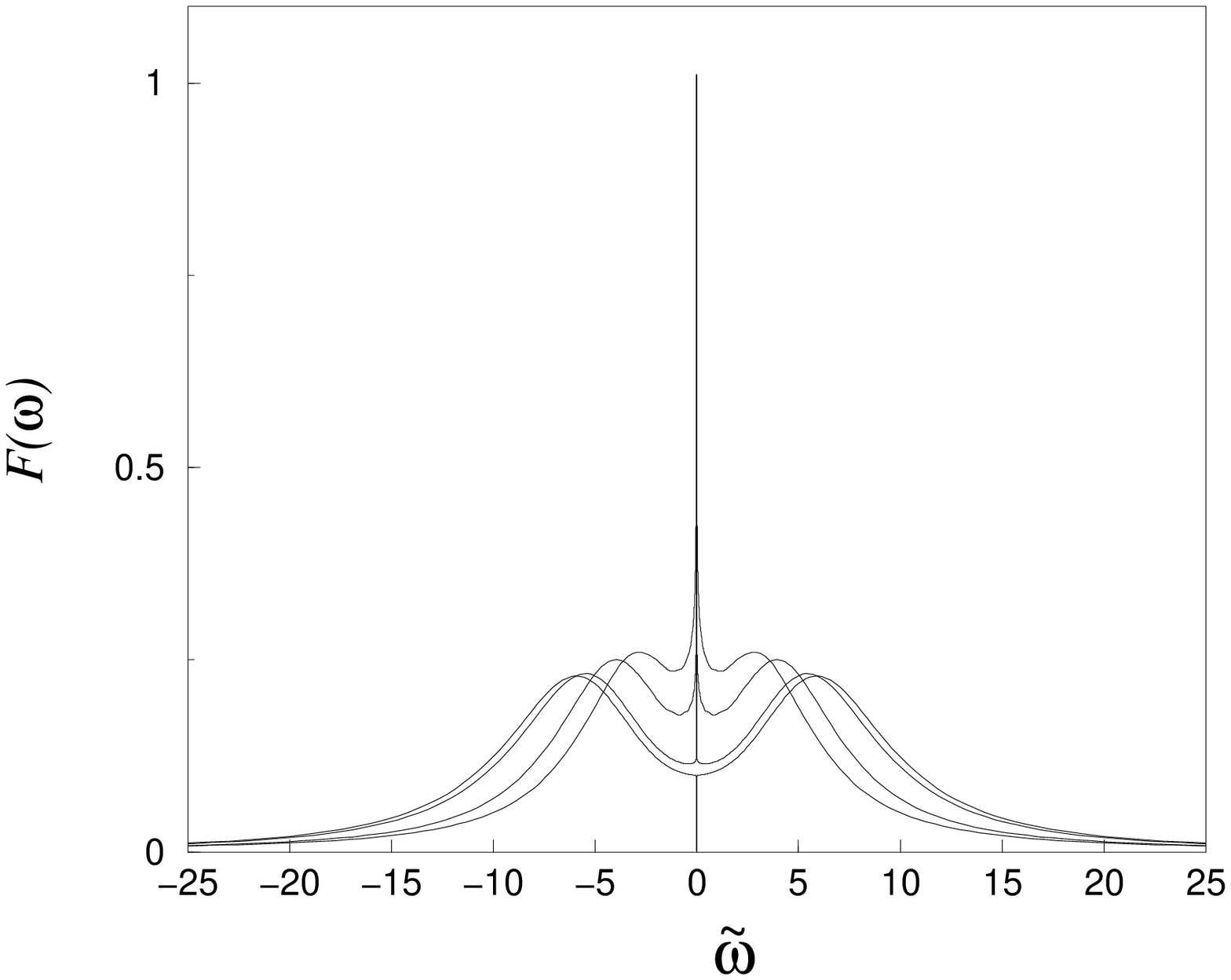}
\caption{NRG spectra ${\cal F}(\w)=\pi\Delta_0[1+\mbox{tan}^2(\frac{\pi}{2}r)]|\w|^rD(\w)$ versus $\tilde{\w}=\w/\delr$ for fixed $r=0.2$ and $U/D=0.1$, and (from inside to outside) for: $\tilde{U}=U/\delr$ = 8.0 (SC), 10.1 (SC), 13.3 (SC) and 14.2 (LM). The SC/LM transition occurs at $\tilde{U}_{\cc}=13.7$.}
\label{fw}
\end{center}
\end{figure}

Note first from Fig.\  9 that the spectral sum rule ${\cal F}(\w=0)=1$ (see \S2), that generalizes to the SC state for arbitrary $r<\frac{1}{2}$ [16] the pinning condition familiar for the $r=0$ Anderson model, and encompasses the latter as a particular case, is well satisfied in the NRG calculations; being obeyed typically to within an accuracy of 1--2\% throughout the SC phase.

As evident from Fig.\  9, and in direct parallel to the normal Anderson model, ${\cal F}(\w)$ in the SC phase contains both the expected Hubbard satellites and a generalized Kondo resonance that narrows progressively as $\tilde{U}$ is increased towards the SC/LM phase boundary at $\tilde{U}_{\cc}(r)$; in fact the SC ${\cal F}(\w)$ shown in Fig.\  9 is qualitatively  indistinguishable from that corresponding to the $r=0$ case on the `low-resolution' scale shown (see e.g.\ Fig.\  6 of [16]).  For concreteness in what follows we define the Kondo scale $\w_{\mbox{\ssz{K}}}\equiv \w_{\mbox{\ssz{K}}}(r)$ as the $\frac{1}{2}$-width at $\frac{1}{2}$-height of ${\cal F}(\w)$.  The definition is of course  somewhat arbitrary, the important point being that $\w_{\mbox{\ssz{K}}}(r)$ should vanish as $\tilde{U}\ra \tilde{U}_{\cc}(r)-$, and within the NRG be proportional to the low-energy scale $\w_*(r)$ (with which it is hence in effect synonymous). Likewise, within the LMA, $\w_{\mbox{\ssz{K}}}(r)$ is proportional to the spin-flip scale $\w_{\m}(r)$ (\S2), and thus correspondingly synonymous.

In contrast to the normal Anderson model, the SC $\w_{\mbox{\ssz{K}}}(r)$ vanishes at a {\it finite} $\tilde{U}_{\cc}(r)$, the effects of which are seen clearly in Fig.\  9 where for $\tilde{U}>\tilde{U}_{\cc}(r)(\simeq 13.7)$ in the LM phase the Kondo resonance is absent, and as $\w \ra 0$ the LM ${\cal F}(\w)\propto |\w|^rD(\w)$ vanishes with characteristic $|\w|^{2r}$ behaviour.  The $\tilde{U}$=13.3 (SC) and 14.2 (LM) examples included in Fig.\  9, corresponding to $\delta=|\tilde{U}/\tilde{U}_{\cc}(r)-1|\simeq 0.03$, also exemplify spectral evolution as the transition is approached from either side.  They show, as argued within the LMA in I (see Fig.\  12 therein), that the SC and LM spectra on either side of the SC/LM transition are near coincident on all energy scales save the lowest; and as $\delta \ra 0$ the LM/SC spectra coincide to arbitrarily low energies, in which sense the spectra evolve smoothly as the phase boundary is approached.

\seceq
\section{Spectra: universal scaling}
Since the Kondo scale $\w_{\K}(r)$ --- and hence the width of the Kondo resonance in ${\cal F}(\w)$ --- vanishes as the SC/LM transition is approached from the SC side, one anticipates spectral scaling, with an $r$-dependent family of universal scaling spectra ${\cal F}\equiv {\cal F}(\w/\w_{\K}(r))$ obtained as $\tilde{U}\ra \tilde{U}_{\cc}(r)-$.  This is indeed predicted by the LMA (\S8 of I), and is now investigated via the NRG.

\begin{figure}[htb]
\begin{center}
\epsfxsize=3.5in
  \epsffile{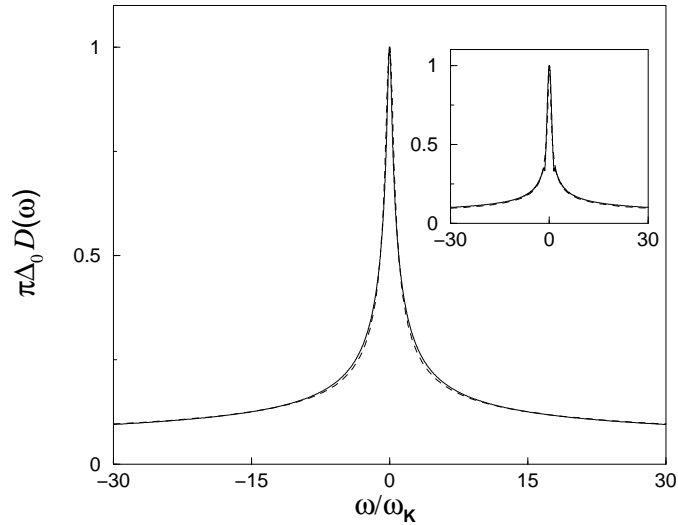}
\caption{For normal Anderson model ($r=0$), NRG scaling spectrum ${\cal F}(\w)=\pi \Delta_0 D(\w)$ versus $\w/\w_{\K}$ (dashed line), compared to that obtained via the LMA (solid line) as discussed in text.}
\label{fig6}
\end{center}
\end{figure}

To this end we first consider the normal Anderson model, $r=0$, where ${\cal F}(\w)=\pi \Delta_0 D(\w)$.  Here, spectral scaling has hitherto been examined via NRG for an asymmetric model [28], in the one-hole sector $\w<0$; and for the symmetric model by a QMC/maximum-entropy study [30] at finite temperatures and for moderate interaction strengths $\tilde{U}=U/\Delta_0 \lesssim 3\pi$.  NRG results for the $T=0$ symmetric Anderson model, ${\cal F}(\w)$ versus $\w/\w_{\K}$, are shown in Figs.\  10 and 11; they were obtained using $\Lambda=1.5$ and $N_{\mbox{\ssz{s}}}=900$, and universal scaling in terms of $\w/\w_{\K}$ is well established for $\tilde{U} \gtrsim 5\pi$.  Fig.\  10 also compares the NRG scaling spectrum to that obtained previously [32] within the LMA.  The inset shows the LMA results (Fig.\  8 of [32]) arising when a particle-hole ladder sum is employed for the transverse spin polarization propagator that enters the LMA self-energies $\tilde{\Sigma}_{\sigma}(\w)$; while the main figure  shows a simple modification detailed in [32] that merely eliminates the small spectral artifact apparent in the former at $\w/\w_{\K} \sim 1.4$, but otherwise produces no significant effect on the scaling spectrum.

\begin{figure}[htb]
\begin{center}
\epsfxsize=3.5in
  \epsffile{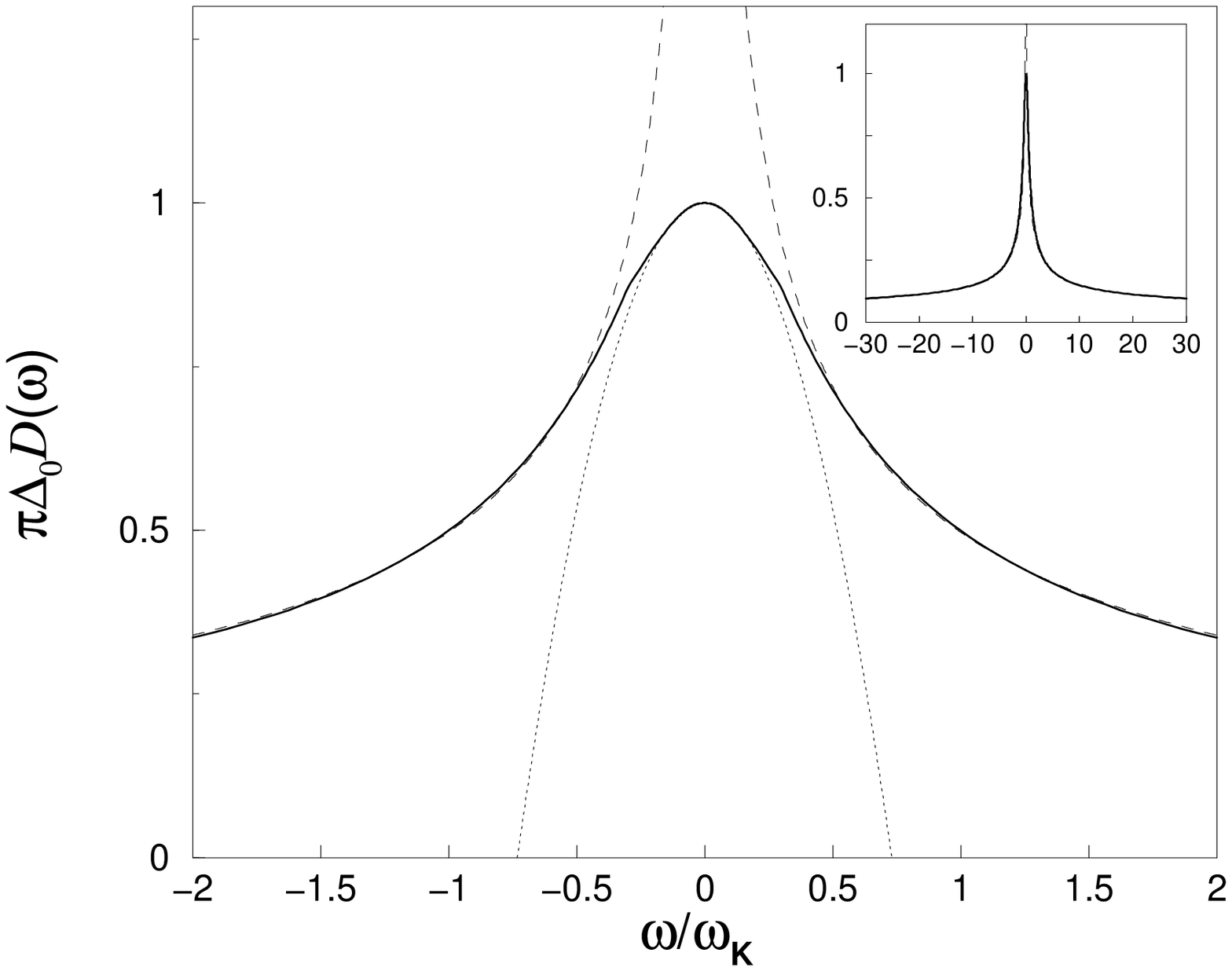}
\caption{For $r=0$ Anderson model, NRG ${\cal F}(\w)=\pi \Delta_0 D(\w)$ versus $\w/\w_{\K}$ including fits to the DS tails (Eq.\  (6.1), dashed line) and to the Fermi liquid portion of the spectrum ($D(\w)-D(0)\propto (\w/\w_{\K})^2$, dotted line).  The inset shows the continued persistence of the DS fit.} 
\label{fit} 
\end{center}
\end{figure}
\begin{figure}[htb]
\begin{center}
\epsfxsize=3.5in
  \epsffile{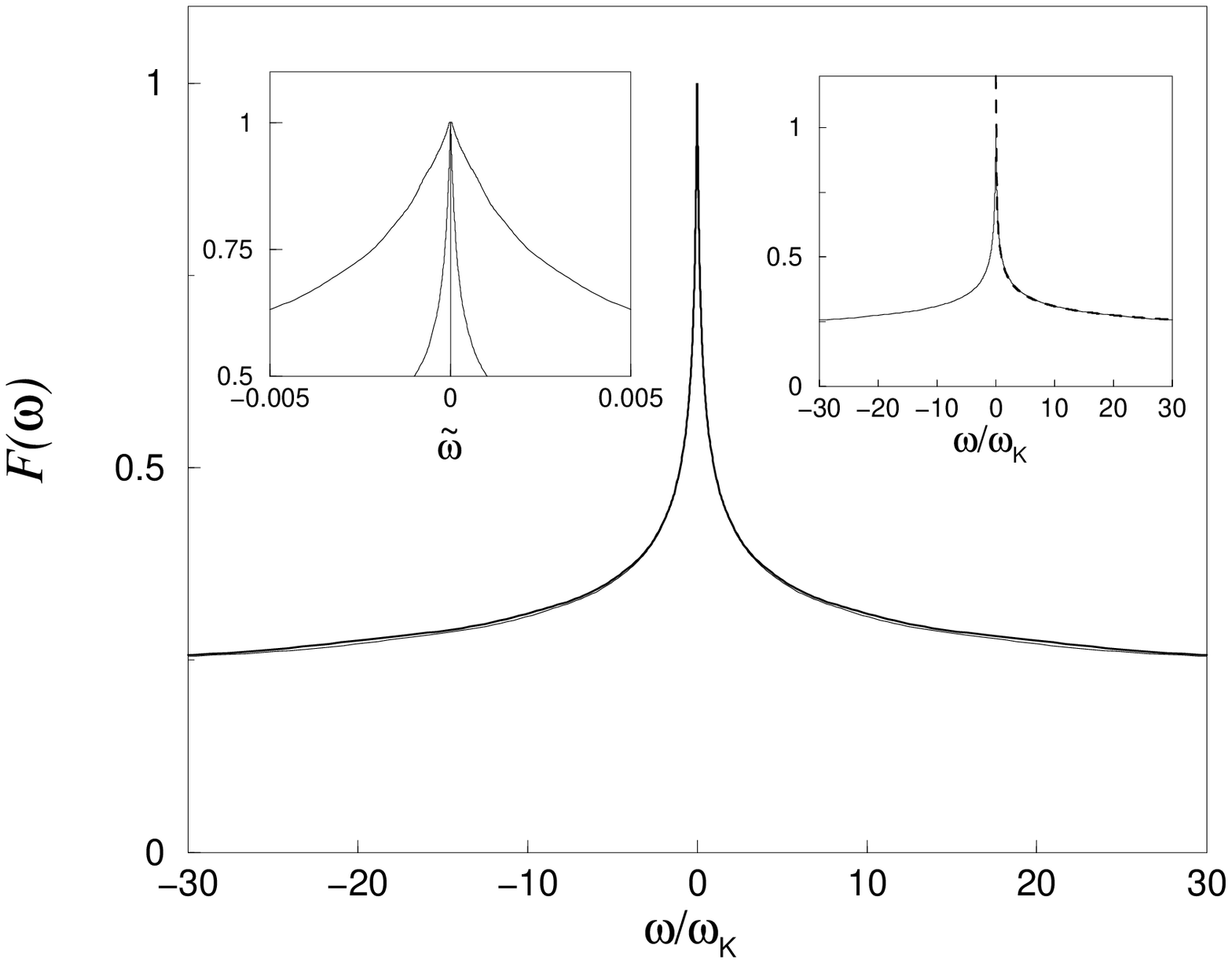}
\caption{For $r=0.2$, NRG scaling spectrum ${\cal F}(\w)=\pi\Delta_0\left[1+\mbox{tan}^2(\frac{\pi}{2}r)\right]|\w|^rD(\w)$ versus $\w/\w_{\K}$; with fixed $U/D=0.1$ and for $\tilde{U}=U/\delr$ = 8.0, 10.1 and 13.3 (the critical $\tilde{U}_{\cc}(r)=13.7$). Left inset: corresponding data versus $\tilde{\w}=\w/\delr$, to illustrate vanishing of Kondo scale as $\tilde{U}\ra \tilde{U}_{\cc}(r)-$. Right inset: DS fit to the scaling spectrum (dashed line), Eq.\  (6.2) with tail exponent $\nu(r)=\frac{1}{2}-r$.}
\label{fig12}
\end{center}
\end{figure}

As seen from Fig.\  10, the agreement between the LMA and NRG scaling spectra for $r=0$ is rather remarkable over the full frequency range.  In addition to spectral pinning at the Fermi level ($\w=0$), the LMA successfully captures both the characteristic low-frequency Fermi liquid behaviour $D(\w)-D(0)\sim (\w/\w_{\K})^2$ and, at larger $\w/\w_{\K}$, the Doniach-\u{S}unji\'{c} (DS) law [27,28,30] indicative of the orthogonality catastrophe, whereby $D(\w)\sim (|\w|/\w_{\K})^{-\alpha}$ with  $\alpha=1-2(\delta_0/\pi)^2$ and $\delta_0=\pi/2$ the Fermi level phase shift; ie $D(\w)\sim (|\w|/\w_{\K})^{-\frac{1}{2}}$.  The existence and, indeed, dominance of the latter is apparent in Fig.\  11, where a DS fit of form
\be
{\cal F(\w)}=a+b(|\w|/\w_{\K})^{-\frac{1}{2}}
\ee 
is made to the NRG data (a fit to the LMA is essentially equivalent).  As seen from the inset to Fig.\  11, DS behaviour persists out to $\w/\w_{\K}=30$ (and naturally well beyond), no deviation from the form Eq.\  (6.1) being apparent; while the main part of Fig.\  11 shows clear DS tail behaviour to extend down to $\w/\w_{\K} \simeq 0.5$.  By contrast, the low-frequency Fermi liquid behaviour persists only {\it up} to $\w/\w_{\K} \simeq 0.2$, where a rapid crossover to DS behaviour begins; this too is seen in Fig.\  11 where extrapolation of the Fermi liquid portion of the spectrum ($D(\w)-D(0)\propto [\w/\w_{\K}]^2$) is made, producing a Lorentzian ${\cal F}(\w)$ that, save for $\w/\w_{\K}\lesssim 0.2$, fails entirely to capture the scaling spectrum.

\begin{figure}[htb]
\begin{center}
\epsfxsize=3.5in
\epsffile{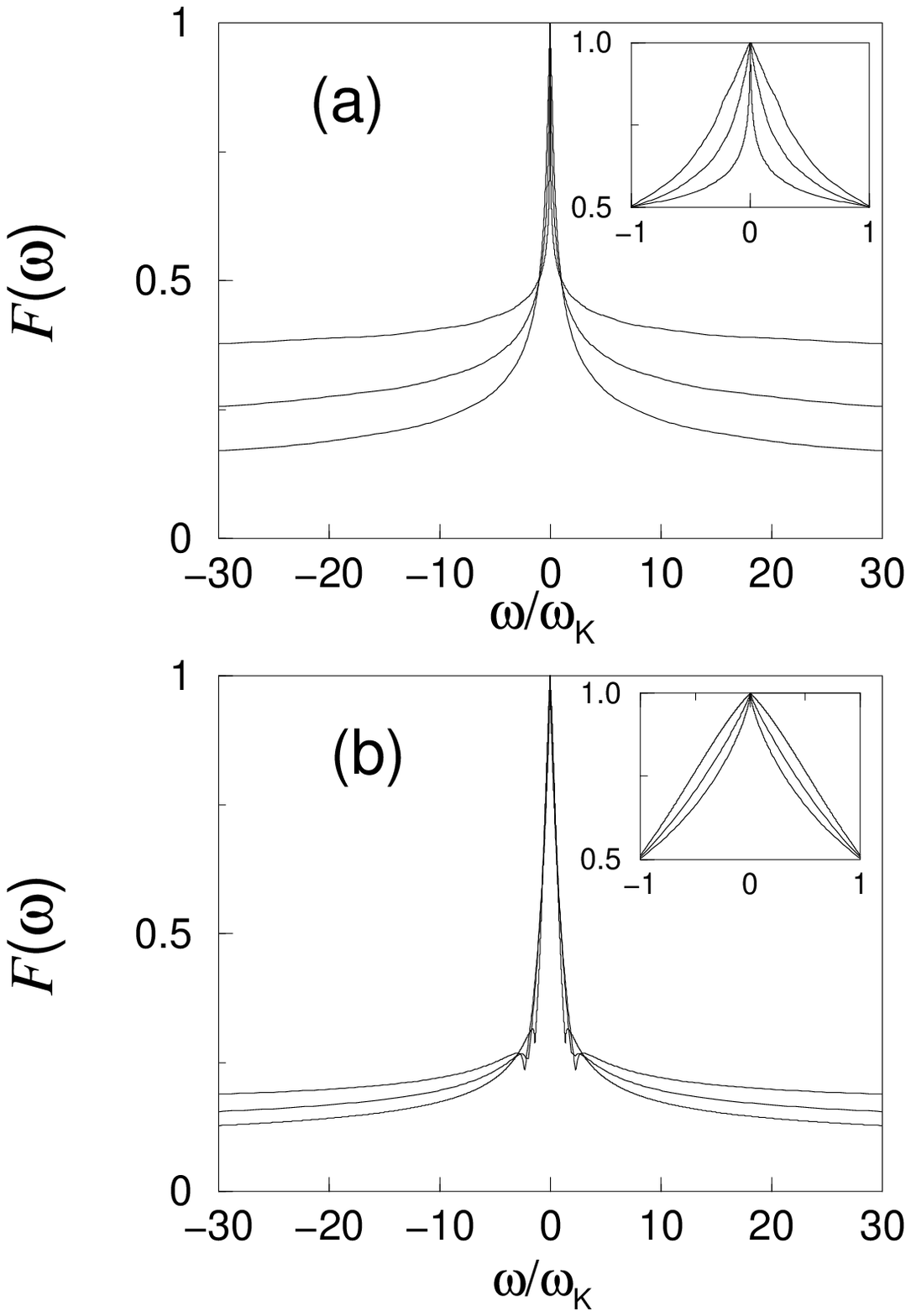}
\caption{Scaling curves ${\cal F}(\w)=\pi\Delta_0\left[1+\mbox{tan}^2(\frac{\pi}{2}r)\right]|\w|^rD(\w)$ versus $\w/\w_{\K}(r)$ for (from bottom to top) $r$ = 0.1, 0.2 and 0.3; as obtained via (a) NRG, and (b) LMA.  Insets: same data on a reduced scale; from outside to inside the curves correspond to $r$ = 0.1, 0.2 and 0.3.}
\label{fig13}
\end{center}
\end{figure}

We turn now to $r>0$, to consider scaling of the SC ${\cal F}(\w)$ as the SC/LM transition is approached.  Specific NRG results shown below are obtained for fixed $U/D=0.1$ (as in \S5); but we note that while the particular value of $\tilde{U}_{\cc}(r)(=(U/\delr)_{\cc})$ at which the transition occurs naturally depends upon $U/D$ (see e.g.\ Fig.\  6), variation of this parameter produces no detectable influence on the scaling spectra themselves.  For $r=0.2$, Fig.\  12 shows the resultant NRG ${\cal F}(\w)$ for $\tilde{U}$ = 8.0, 10.1 and 13.3 (as in Fig.\  9); the critical $\tilde{U}_{\cc}(r)\simeq 13.7$.  
The left-hand inset to the Fig.\  shows the central portion of the Kondo resonance, versus $\tilde{\w}=\w/\delr$ on an `absolute' scale, to illustrate its rapid narrowing with increasing $\tilde{U}$, and ultimate vanishing as $\tilde{U}\ra \tilde{U}_{\cc}-$; the same behaviour is clearly evident within the LMA in Fig.\  18 (inset) of I.  By contrast, the main part of Fig.\  12 shows the scaled ${\cal F}(\w)$ versus $\w/\w_{\K}$, from which universality is indeed apparent; the scaled spectra for $\tilde{U}$ =10.1 and 13.3 in particular are to all intents and purposes indistinguishable.  Note also that while $\tilde{U}_{\cc}(r)$ is finite, the Hubbard satellites are naturally projected out of the scaling spectrum since $\w_{\K}(r)$ vanishes at the phase boundary.

\begin{figure}[htb]
\begin{center}
\epsfxsize=3.5in
  \epsffile{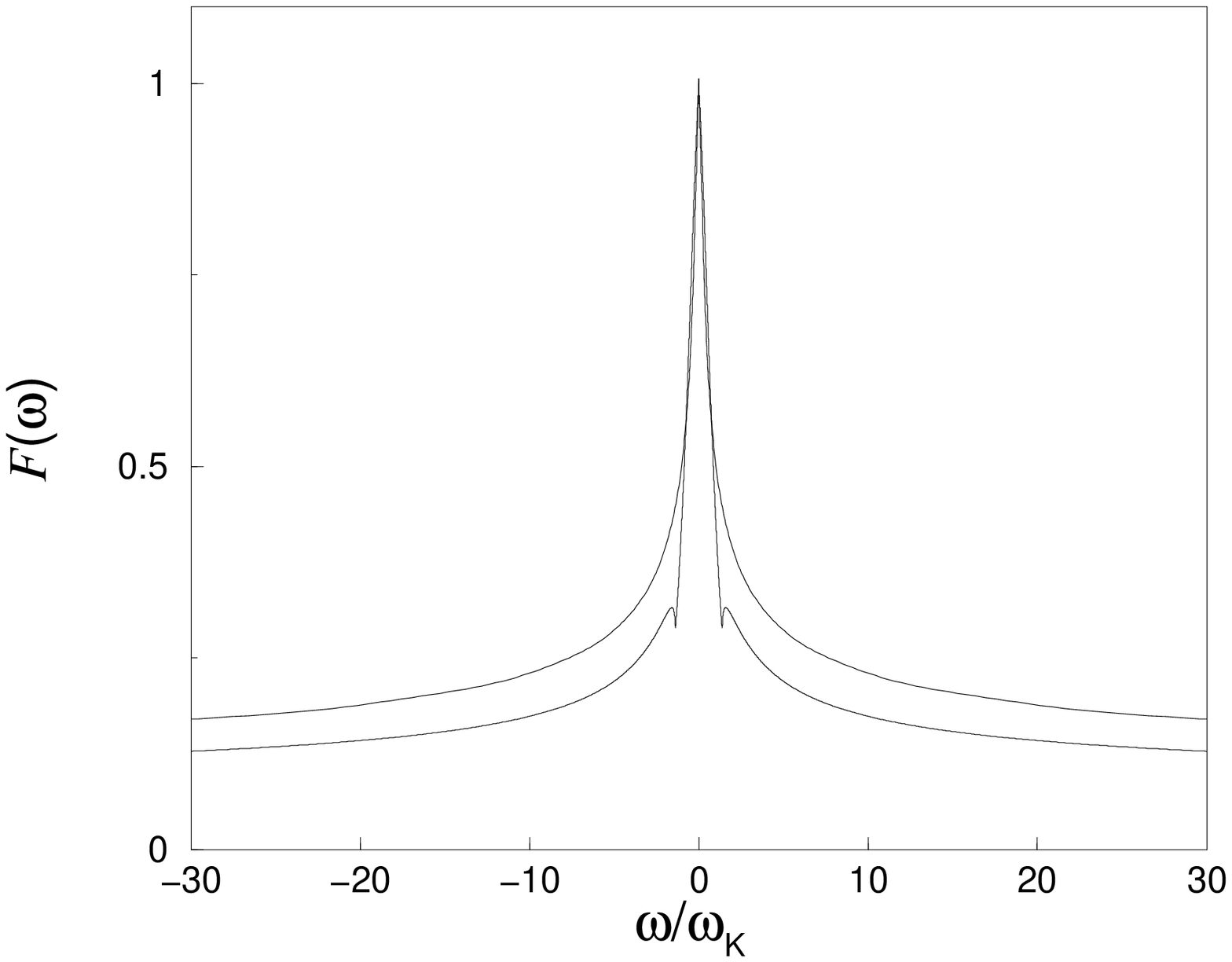}
\caption[]{Direct comparison of $r=0.1$ scaling curves ${\cal F}(\w)=\pi\Delta_0\left[1+\mbox{tan}^2(\frac{\pi}{2}r)\right]|\w|^rD(\w)$ versus $\w/\w_{\K}$: NRG (upper) and LMA (lower).}
\label{fig14}
\end{center}
\end{figure}

The LMA predicts two essential characteristics of the SC spectra (\S8 of I).  First, that for $r>0$ the leading $\w/\w_{\K}(r) \ra 0$ behaviour of ${\cal F}(\w)-1 (={\cal F}(\w)-{\cal F}(0))$ is ${\cal F}(\w)-1 \propto [|\w|/\w_{\K}]^{\gamma(r)}$ with $\gamma(r)=1-r$ (and a proportionality constant that naturally vanishes as $r \ra 0$, see \S8 of I).  Cusp behaviour is indeed seen in the $r=0.2$ NRG results of Fig.\  12, and numerical analysis confirms an exponent of $\gamma(r)=0.8$ (asymptotic behaviour that in practice sets in for $|\w|/\w_{\K}\lesssim 0.01$).  Second, that the wings of the scaling spectrum should contain generalized DS tails, of form ${\cal F}(\w)\sim [|\w|/\w_{\K}(r)]^{-\nu (r)}$ with an exponent $\nu (r)=\frac{1}{2}-r$.  This is likewise found in the NRG calculations, as seen from the right-hand inset to Fig.\  12 for $r=0.2$, where a DS fit of form
\be
{\cal F}(\w)=a+b[|\w|/\w_{\K}(r)]^{-(\frac{1}{2}-r)}
\ee
is made.  The fit is excellent (extending down to $|\w|/\w_{\K}\sim 0.5$) and, as for the $r=0$ model, the DS tails are again seen to dominate the scaling spectrum.  The above behaviour is not of course confined to $r=0.2$.  
Fig.\  13a shows NRG results for ${\cal F}(\w)$ versus $\w/\w_{\K}(r)$, for $r$ = 0.1, 0.2 and 0.3 (from bottom to top); the inset shows the same data on a reduced scale $|\w|/\w_{\K}(r)<1$.  Corresponding LMA results are shown in Fig.\  13b (obtained using the particle hole ladder sum in $\tilde{\Sigma}_{\sigma}(\w)$, and with no attempt to avoid the spectral anomaly apparent around $|\w|/\w_{\K}\sim$ 1--2).  In all these cases, for both NRG and LMA spectra, the cusp exponent is found to be $\gamma(r)=1-r$; and DS tails of form Eq.\  (6.2) again arise.  We also add that the DS exponent of $\nu(r)=\frac{1}{2}-r$ reflects the fact that we consider ${\cal F}(\w)\propto |\w|^rD(\w)$ which, unlike $D(\w)$ itself, is a universal function of $\w/\w_{\K}(r)$ as the transition is approached from the SC phase.  However $[\w_{\K}(r)]^rD(\w)\propto (|\w|/\w_{\K}(r))^{-r}{\cal F}(\w)$ does exhibit scaling, and we note that its DS exponent is thus $\nu(r)+r=\frac{1}{2}$ --- precisely as for the normal Anderson model, and again symptomatic of the fact that the latter consitutes a particular case of more generic SC behaviour.

In qualitative terms, both NRG and LMA results for the universal ${\cal F}(w)$ in Fig.\  13 show narrowing of the central portion of the Kondo resonance with increasing $r$ (insets to Fig.\  13); together with a relative `flattening' ($\nu(r)=\frac{1}{2}-r$) and increase in intensity of the DS tails.  Quantitatively, the agreement between NRG and LMA for ${\cal F}(\w)$ becomes poorer with increasing $r$ --- not surprisingly in view of the discussion of the phase boundaries given in \S4, and reflecting the fact that the specific LMA we have considered focuses on the spin-fluctuation physics which is asymptotically dominant as $r\ra 0$.  
Explicit comparison between NRG and LMA results for the universal ${\cal F}(\w)$ is made in Fig.\  14 for $r=0.1$; the most obvious difference being the relative offset in the DS tails (each of which fits very well to the form Eq.\  (6.2)), which increases with $r$ (Fig.\  13).  This shows up in significant differences between the NRG and LMA $a$-coefficients in Eq.\  (6.2); although the corresponding $b$-coefficients are much closer, reflected in the fact that the DS tails in Fig.\ 14 are near parallel.

\section{Summary}
In this paper we have studied the symmetric soft-gap Anderson impurity model
[9,14-16,31,32]. The model exhibits a rich spectrum of physical behaviour
on account of the underlying quantum phase transition between strong coupling
and local moment states, and contains the `normal' Anderson model as
a particular limiting case. Our primary aim has been to make detailed
comparison between two non-perturbative approaches: the NRG method, which
provides essentially exact numerical results for the problem [14,15],
and a recently developed many-body theory, the LMA [31,32].

  We find excellent agreement between the two approaches,
both for static properties such as the ground state phase diagram, as
well as for dynamical properties embodied in zero temperature
single-particle excitation spectra. In particular, specific
predictions arising from the LMA [31] are found to be well borne out by NRG results.
These include scaling characteristics and asymptotic behaviour of the
phase boundaries, Fermi level pinning of the modified spectral functions
throughout the SC phase, and universal scaling of single-particle spectra 
upon approach to the SC$\rightarrow$LM transition where the Kondo
scale characteristic of the SC phase vanishes.  Differences
between the two approaches arise only as $r\rightarrow \frac{1}{2}$; they may be ascribed
to an incomplete treatment of charge fluctuations in the specific LMA we
have considered in practice, which seeks primarily to capture the spin-flip
physics that  dominates the regime of well-developed local moments (whether in the
SC or LM phases). Quantitative agreement between NRG and LMA results is however
remarkably good in general, the universal scaling spectrum of the normal Anderson
model (Fig.10) providing a particular example. The results obtained here provide
in our view strong encouragement to further study of the present ---and related---
quantum impurity models, via both the NRG and LMA, and including consideration of
finite-temperature spectral and hence transport properties.

\ack
We are grateful to Alex Hewson for helpful discussions, and to the British Council and DAAD for financial support via an ARC grant.  MTG gratefully acknowledges the award of an EPSRC studentship.

\end{document}